\pdfoutput=1
\documentclass[final,5p,times,twocolumn]{elsarticle}
\usepackage{amssymb}
\usepackage{dcolumn}
\usepackage{bm}

\def\pbar{\bar{p}}

\def\Hbar{\bar{\mathrm{H}}}

\begin{document}

\begin{frontmatter}

\title{
Monte-Carlo based performance assessment of ASACUSA's antihydrogen detector
}

\author[myu,myu2]{Y. Nagata\corref{mycorrespondingauthor}}
\cortext[mycorrespondingauthor]{Corresponding author}
\ead{yugo.nagata@rs.tus.ac.jp}

\address[myu]{Department of Physics, Tokyo University of Science, 1-3 Kagurazaka, Shinjuku, 162-8601 Tokyo, Japan}
\address[myu2]{Ulmer Fundamental Symmetries Laboratory, RIKEN, 2-1 Hirosawa, Wako-shi, 351-0198 Saitama, Japan}
\address[myu3]{Institute of Physics, University of Tokyo, 3-8-1 Komaba, Meguro-ku, 153-8902 Tokyo, Japan}
\address[myu4]{Stefan-Meyer-Institut f$\ddot{u}$r Subatomare Physik, $\ddot{O}$sterreichische Akademie der Wissenschaften, Wien 1090, Austria}
\address[myu5]{CERN, Gen$\grave{e}$ve 1211, Switzerland}
\address[myu6]{Graduate School of Advanced Sciences of Matter, Hiroshima University, 739-8530 Hiroshima, Japan}
\address[myu7]{RIKEN Nishina Center for Accelerator-Based Science, 2-1 Hirosawa, Wako-shi, 351-0198 Saitama, Japan}
\address[myu8]{Dipartimento di Ingegneria dell'Informazione, Universit$\grave{a}$ di Brescia, Brescia 25133, Italy}
\address[myu9]{Istituto Nazionale di Fisica Nucleare, Sez. di Pavia, I-27100 Pavia, Italy}

\author[myu3]{N. Kuroda}
\author[myu4]{B. Kolbinger}

\author[myu4]{M. Fleck\fnref{pad}}
\fntext[pad]{Present address: Institute of Physics, University of Tokyo, 3-8-1 Komaba, Meguro-ku, 153-8902 Tokyo, Japan}

\author[myu4,myu5]{C. Malbrunot}

\author[myu4]{V. M$\mathrm{\ddot{a}}$ckel\fnref{pad2}}
\fntext[pad2]{Present address: Ulmer Fundamental Symmetries Laboratory, RIKEN, 2-1 Hirosawa, Wako-shi, 351-0198 Saitama, Japan}

\author[myu4]{C. Sauerzopf\fnref{pad4}}
\fntext[pad4]{Present address: Data Technology, Vienna, Austria}

\author[myu4]{M. C. Simon}
\author[myu3]{M. Tajima\fnref{pad3}}
\fntext[pad3]{Present address: RIKEN Nishina Center for Accelerator-Based Science, 2-1 Hirosawa, Wako-shi, 351-0198 Saitama, Japan}

\author[myu4]{J. Zmeskal}
\author[myu5]{H. Breuker}
\author[myu6]{H. Higaki}
\author[myu7]{Y. Kanai}
\author[myu3]{Y. Matsuda}
\author[myu2]{S. Ulmer}
\author[myu8,myu9]{L. Venturelli}
\author[myu4]{E. Widmann}
\author[myu2]{Y. Yamazaki}

\date{\today}

\begin{abstract}
An antihydrogen detector consisting of a thin BGO disk and a surrounding plastic scintillator hodoscope has been developed. We have characterized the two-dimensional positions sensitivity of the thin BGO disk and energy deposition into the BGO was calibrated using cosmic rays by comparing experimental data with Monte-Carlo simulations. The particle tracks were defined by connecting BGO hit positions and hits on the surrounding hodoscope scintillator bars. The event rate was investigated as a function of the angles between the tracks and the energy deposition in the BGO for simulated antiproton events, and for measured and simulated cosmic ray events. Identification of the antihydrogen Monte Carlo events was performed using the energy deposited in the BGO and the particle tracks. The cosmic ray background was limited to 12~mHz with a detection efficiency of 81~\%. The signal-to-noise ratio was improved from 0.22~${\mathrm{s^{-1/2}}}$
obtained with the detector in 2012 to 0.26~${\mathrm{s^{-1/2}}}$ in this work.

\end{abstract}


\end{frontmatter}

\section{\label{sec:level1}Introduction}
Recently, antihydrogen ($\Hbar$) atoms have been produced \cite{ASACUSA} in a unique cusp trap \cite{MY,YY,DC} developed for the in-flight hyperfine spectroscopy of ground state $\Hbar$ atoms \cite{SMI,HFS1,HFS2}. The most recent progress is reported in \cite{NK,MT,chloe}. In 2012, the ASACUSA Cusp collaboration developed a $\Hbar$ detector consisting of a BGO ($\mathrm{Bi_4Ge_{3}O_{12}}$) scintillator disk in combination with a single anode photomultiplier (PMT) and 5 plastic scintillator plates. The detector was able to reject cosmic backgrounds with a high efficiency \cite{YN1}.
In order to further improve the background rejection efficiency, we have developed a new $\Hbar$ detector. The single anode PMT has been replaced by 4 multi-anode PMTs (MAPMTs) for 2D photon readout of the BGO. The five plastic scintillators were replaced by a two-layer hodoscope with 32 plastic scintillator bars per layer to determine charged particle tracks with higher resolution~\cite{CS2}.\par

\begin{figure}[htb]
\includegraphics[width=1.\linewidth]{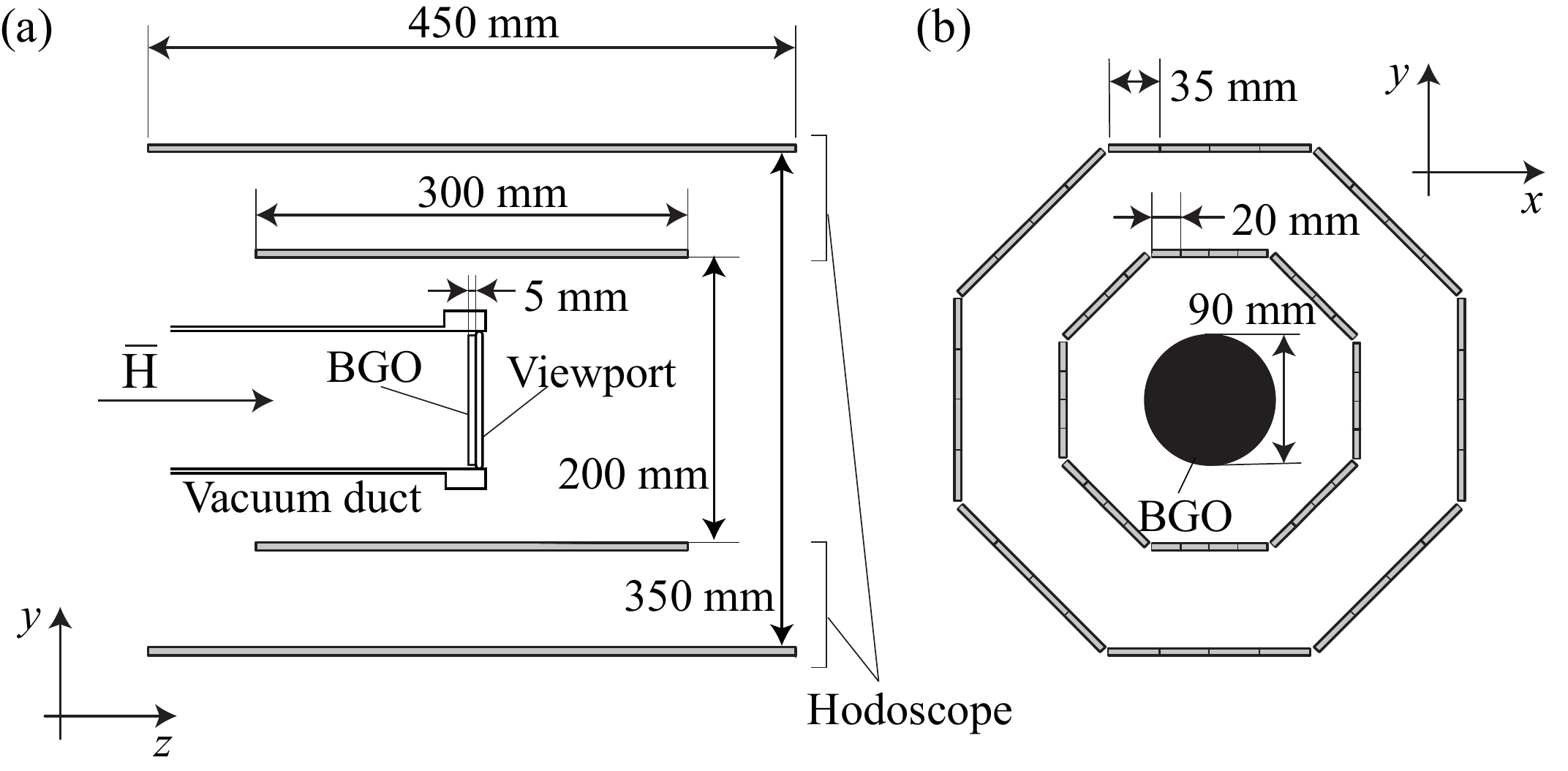}
\caption{\label{fig:detectorsetup}
Cross section of the $\Hbar$ detector along the beam axis (a)
and perpendicular to the $\Hbar$ beam axis (b).
}
\end{figure}

\begin{figure}[htb]
\includegraphics[width=0.8\linewidth]{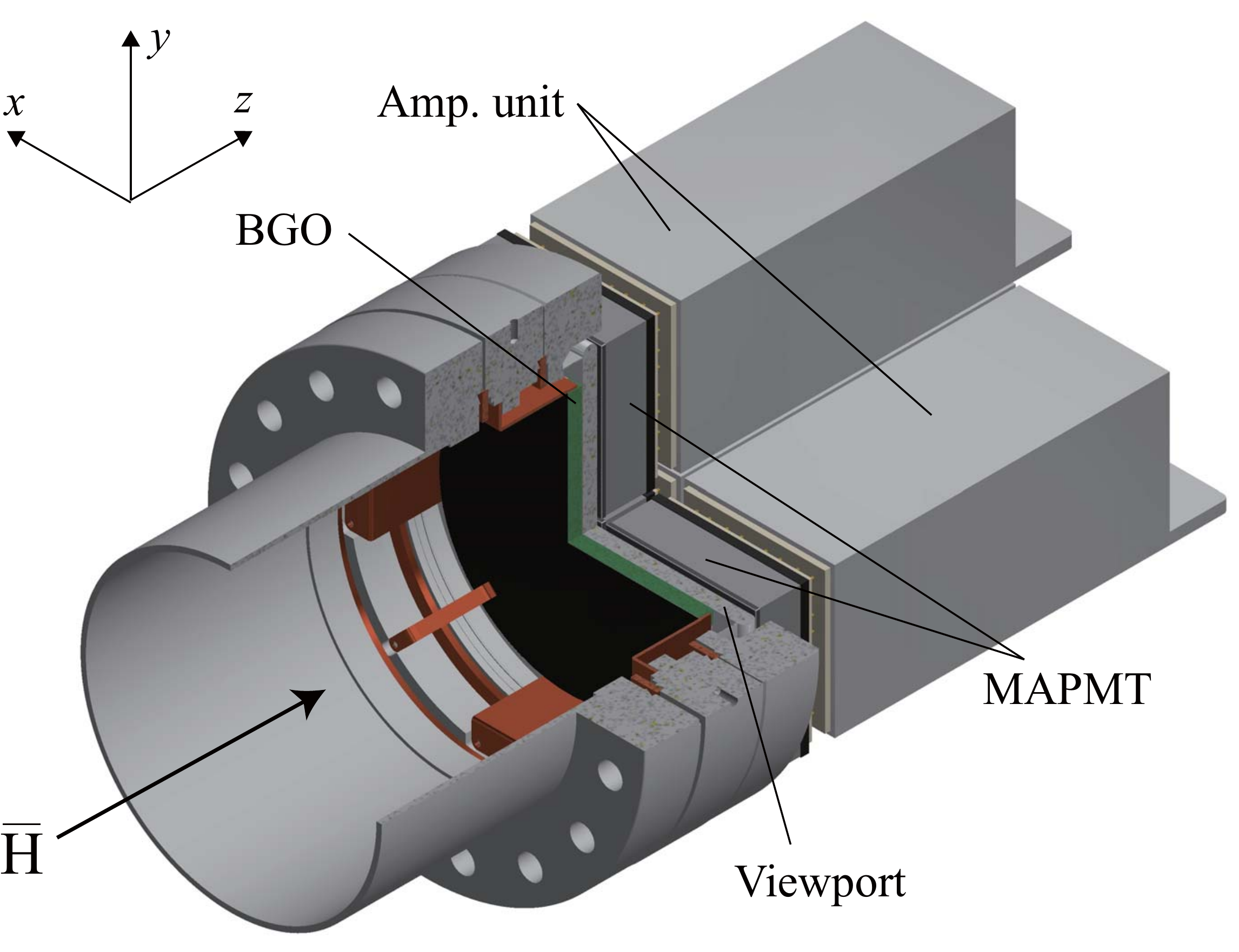}
\caption{\label{fig:BGOPMT1}
Three quarter section view of the 2D sensitive BGO detector.
}
\end{figure}
\section{\label{sec:level2} $\Hbar$ detector}
Figure~{\ref{fig:detectorsetup}} shows a schematic diagram of the structure of the new $\Hbar$ detector consisting of the thin BGO disk and the hodoscope. The BGO disk, has a diameter of 90~mm and  a thickness of 5~mm and is housed on the vacuum side ($10^{-7}$~Pa) of a UHV viewport. The front surface of the BGO disk was coated with a carbon layer of thickness 0.7~$\mu$m to reduce multireflections of the light from scintillation on the surface. It was found that the carbon coating improved the position resolution by a factor of $\sim$~2 in our previous device \cite{YN2}. To achieve a position sensitive readout, 4 MAPMTs (Hamamatsu H8500C) each having 8~$\times$~8 anodes with effective area of 49~mm~$\times$~49~mm were directly placed on the view port glass as shown in Fig.~{\ref{fig:BGOPMT1}}. The output of $8\times 8$ anodes were amplified, digitized and stored by an amplifier unit (Clear Pulse 80190) which was a dedicated model for the H8500C and included $8\times 8$ charge amplifiers and analogue-to-digital converters with 12 bit resolution.\par

The hodoscope consists of two layers of 32 plastic scintillator bars arranged in an octagonal configuration \cite{CS2}. The scintillator bars are 300~mm~$\times$~20~mm~$\times$~5~mm for the inner layer and 450~mm~$\times$~35~mm~$\times$~5~mm for the outer. With face to face distances of 200~mm and 300~mm respectively for the inner and outer layers (see Fig.~{\ref{fig:detectorsetup}}(a)). The solid angle covered by the scintillator bars in units of $4\pi$
seen from the center of the BGO is $\omega \sim 80$~\%. Silicon photomultipliers (SiPM, KETEK PM3350TS) were connected to both ends of each bar. The output pulses from the SiPMs was amplified by dedicated front end modules described in detail elsewhere (see ref \cite{CS1}) and recorded by 128 channels waveform digitizers (CAEN V1742).

\section{\label{sec:level3}2D distribution of cosmic rays}
\begin{figure}[hb]
\includegraphics[width=1.\linewidth]{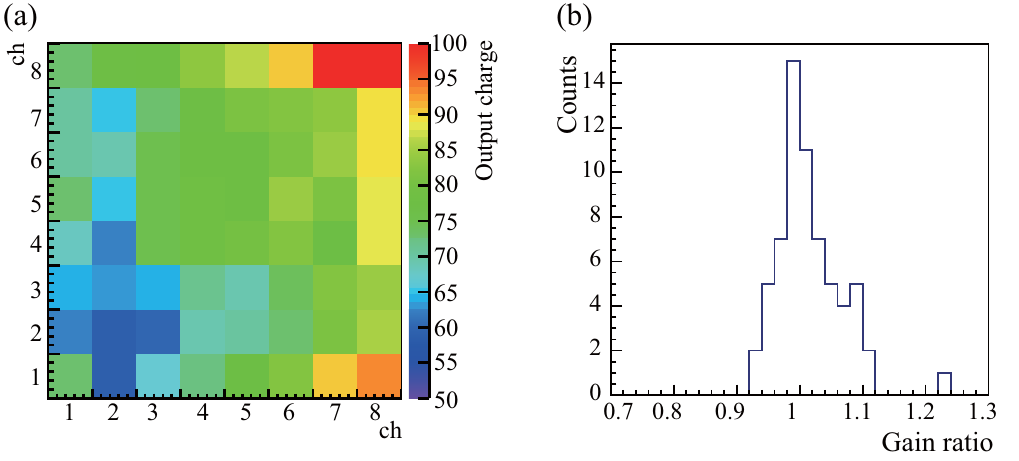}
\caption{\label{fig:compareHamamatsu}
(a) Example of a 2D map of averaged output charges of one of the 4 MAPMTs investigated by LED light.
(b) Distribution of the gain ratio between the measured values and the data sheet values for each channel.
}
\end{figure}

To obtain the relative sensitivity of each channel, 4 MAPMTs were assembled and irradiated by pulsed (200~ns width) light from a LED with the peak wavelength of 470~nm (see also Ref. \cite{YN2}), the peak emission wavelength of the BGO scintillation light was 480~nm.
The voltages applied to the MAPMTs were adjusted such that the total output charge of each MAPMT were equal.

Figure~{\ref{fig:compareHamamatsu}}~(a) shows an example of a 2D map of output charges from one of the 4 MAPMTs averaged over $10^4$ pulses of LED light. The channel with the maximum output charge is arbitrarily set to 100 and all other channels are scaled accordingly. The relative gain of the channels of each MAPMT are evaluated using this mapping. This result was compared with the data sheet from the manufacturer where a tungsten filament lamp was used for calibration. By taking the ratio between the measured values and those from the data sheet for each anode, the distribution of the gain ratio as shown in Fig.~{\ref{fig:compareHamamatsu}}~(b) was obtained. The standard deviation of this distribution is around 5\%.

\begin{figure}[hb]
\includegraphics[width=1.\linewidth]{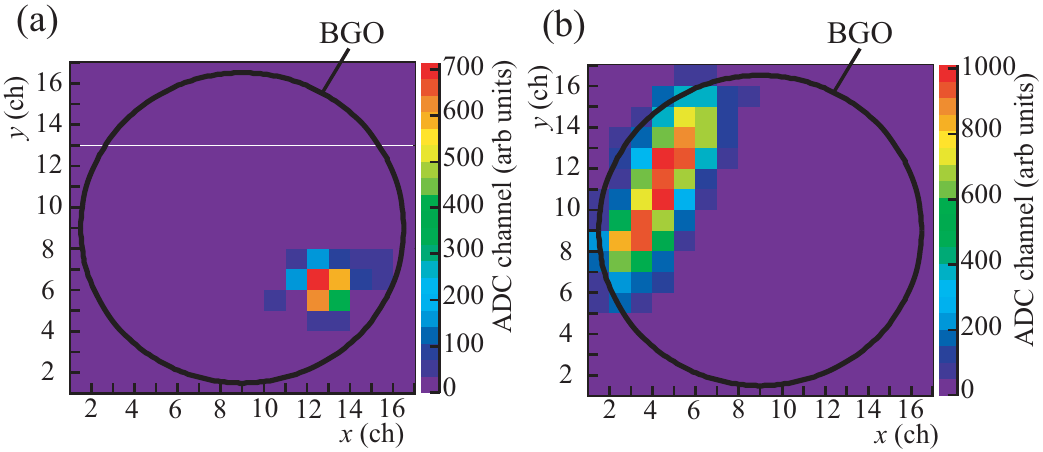}
\caption{\label{fig:cosmicexample}
Examples of 2D charge distributions of cosmic rays events where the BGO surface has been penetrated nearly perpendicular (a) or parallel (b). 
}
\end{figure}

Figures~{\ref{fig:cosmicexample}}~(a) and (b) show the 2D charge distribution for two example cosmic rays events after offline gain matching of the individual MAPMT channels described above. In these examples, the BGO surface has been penetrated nearly perpendicular (a) or parallel (b). These figures demonstrate that position sensitive readout has been successfully implemented using 2D MAPMTs.

In order to reconstruct particle tracks (to be discussed in detail later),
we define the position of a hit on the BGO as the center of the anode giving the highest output charge. When the hit positions observed in the 2D distribution are outside of the BGO, such events are removed in the analysis because they are probably \v{C}erenkov light generated in the glasses of the MAPMT or the viewport when high energy charged particles pass through them. The total charge $Q$ is obtained by summing the charges from all channels (see Figs.~{\ref{fig:cosmicexample}}~(a) and (b)).

\section{\label{sec:level4}Energy calibration of the BGO detector}

\begin{figure}[htb]
\includegraphics[width=0.95\linewidth]{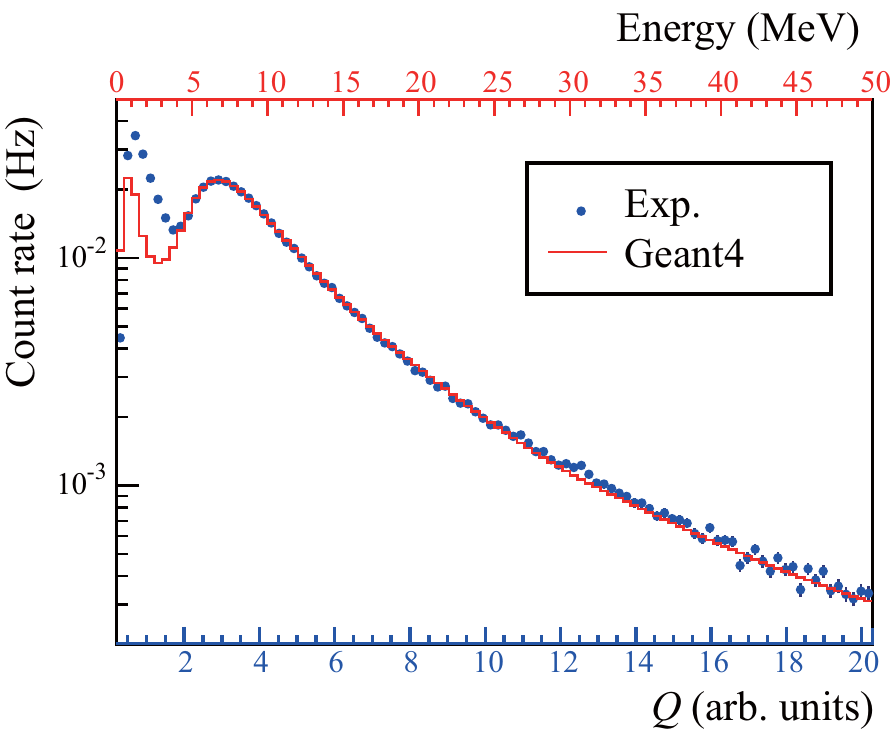}
\caption{\label{fig:chi2Fit}
The blue solid circles show the total measured output charge of cosmic ray events. 
The red line show the simulated energy deposition distributions.
The integral of events with $E\geq50$~MeV was 3 \%
of those with $E\geq4.5$~MeV.
}
\end{figure}

For the energy calibration, cosmic rays were measured and the charge distribution $f(Q)$ was compared with the energy deposition distribution $g(E)$ calculated by a Monte-Carlo simulation using the GEANT4 toolkit\footnote{Geant4.9.6 Patch-02 was used.} \cite{Geant}.
Blue solid circles in Fig.~{\ref{fig:chi2Fit}} show $f(Q)$ for cosmic rays events when more than 2 inner hodoscope bars are hit in any coincident combination. A bar is considered to be 'hit' when there is  a coincidence between the signals of the upstream and downstream SiPMs connected to the bar.

The simulation includes, the BGO detector together with the viewport and the vacuum duct.
Cosmic rays are generated by the CRY package \cite{Hagmann}, however, \v{C}erenkov light is not taken included. $g(E)$ is convoluted with the energy resolution which is assumed to be proportional to $\sqrt{E}$ \cite{leo}, i.e.,
$$
G(E) = \int g(E') \frac{1}{\sqrt{2\pi (\alpha \sqrt{E'})^2}} e^{-(E-E')^2/2 (\alpha \sqrt{E'})^2} dE'\textrm{,}
$$
where $\alpha$ is a fitting parameter.
To compare $f(Q)$ and $G(E)$, the relation $E=\eta Q+E_c$ is assumed, where $\eta$ and $E_c$ are fitting parameters.
$E_c$ corresponds to the contribution of \v{C}erenkov light generated in the BGO and the glass.
$G(E)$ is fit to $f(Q)$ using a least square method in the energy range from 4.5 to 50 MeV with 3 free parameters $\alpha$, $\eta$ and $E_c$.
The result is shown by the red line in Fig.~{\ref{fig:chi2Fit}}.
The goodness of fit is given by $\mathrm{\chi^2/ndf}=1.65$ for $\alpha$=0.52~${\sqrt{{\mathrm{MeV}}}}$, $\eta$=2.5
in units proportional to charge over energy and $E_c$=0.50~MeV.

The cosmic ray flux distribution is known to follow $\cos^2 \theta_z$, where $\theta_z$ is the zenith angle.
In this case, the average path length of cosmic rays in the BGO is evaluated to be approximately 8~mm.
Considering that the energy deposited by a minimum ionizing particle (MIP) in the BGO  is 9.0~MeV/cm \cite{PDG}, the peak at around 7~MeV in Fig.~{\ref{fig:chi2Fit}} is attributed to MIPs. As explained below (see subsection 5.3), only events for $E>E_{th}=15$~MeV are considered in the following sections.

\section{\label{sec:level5}Identification of $\Hbar$ atoms and cosmic ray suppression}
\begin{figure}[t] 
\includegraphics[width=.6\linewidth]{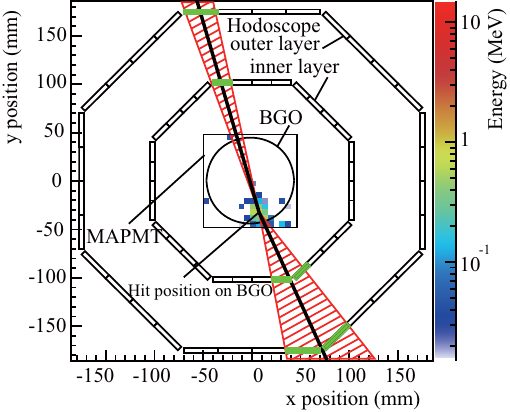}
\caption{\label{fig:tracktheta}
Example of a cosmic ray event with $k=2$ tracks. The circle surrounded by the square in the center show the BGO disk and the total area covered by the 4 MAPMTs, respectively. The external octagonal configuration show the inner and outer hodoscope layers.
}
\end{figure}

Figure~{\ref{fig:tracktheta}} shows a simplified back view of the $\Hbar$ detector where the signals from the 4 MAPMTs (shown in the square in the center) and the hodoscope bars (the outermost octagonal arrangement) for a typical cosmic event in the experiment are overlaid. 

In this example, a spot like pattern is seen on the BGO. Green-colored hodoscope bars are simultaneously hit by the cosmic ray. The red shaded area in the upper left side shows the possible spatial range of the trajectory of the charged particle.

As a best guess, the bisector of the shaded area is taken to define a track which is shown by the thick black line. In the lower right side, 2 neighboring hodoscope bars are hit, in this case, the red shaded area is defined from the edges of the neighboring hodoscope bars and the BGO hit position. The particle track is also defined by a bisector of the shaded area. In this example, the number of tracks is $k=2$.

\begin{figure}[t]
\includegraphics[width=1.\linewidth]{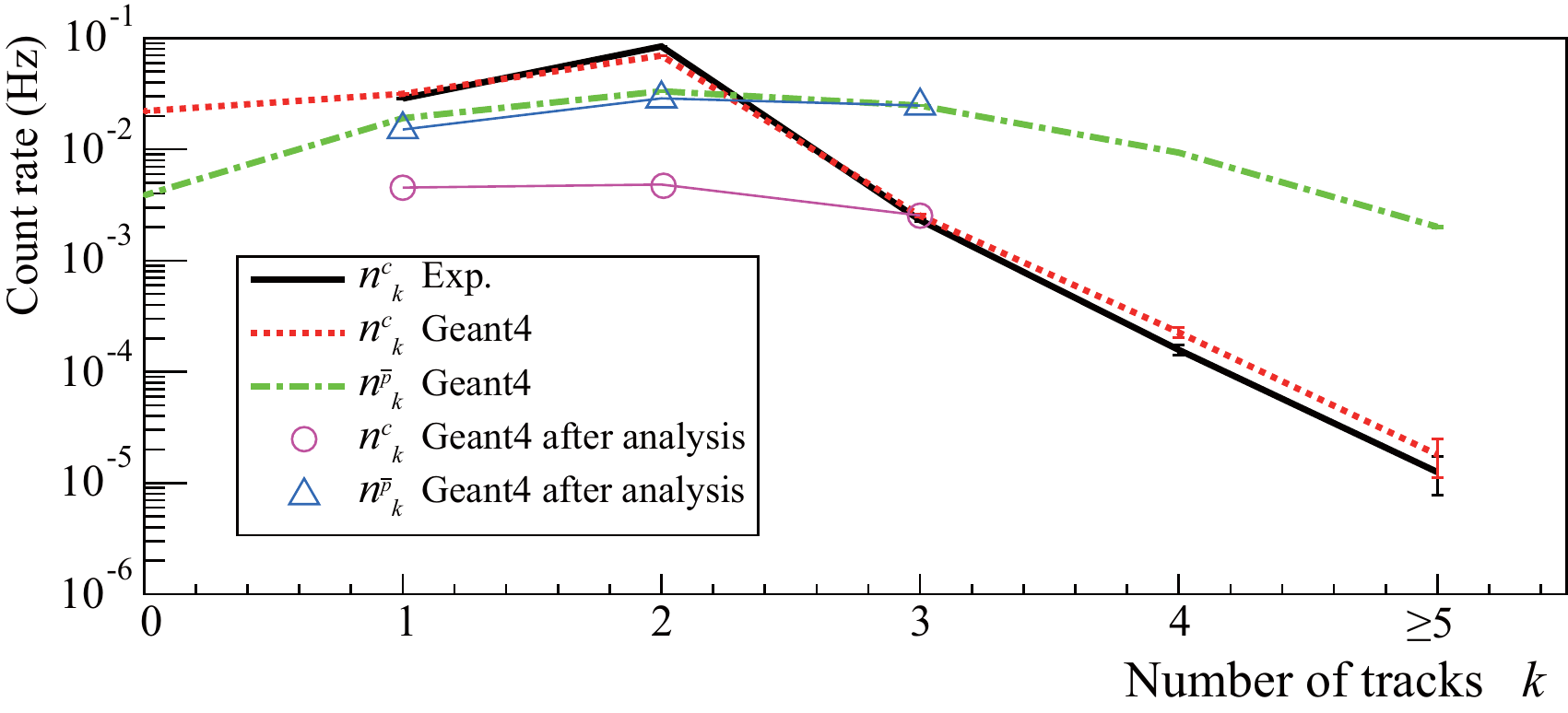}
\caption{\label{fig:Count15MeV45MeV}
Solid and dashed lines show experimental and simulated results of cosmic ray count rate $n^c_k$ as a function of $k$, respectively. The chain curve shows the simulation result of the annihilation count rate $n^{\pbar}_k$ when $\pbar$s irradiate the BGO disk uniformly with a rate of $I_{\pbar}=0.1$~Hz.
Only events depositing more than 15~MeV in the BGO were considered for both cosmics and $\pbar$s in the simulated and experimental data. The open circles and triangles show the simulated data of $n^{c}_k$ and $n^{\pbar}_k$ for $k=1$--3 obtained from the analysis described in the subsection~{\ref{sec:level54}}, respectively.
}
\end{figure}

Solid and dashed lines in Fig.~{\ref{fig:Count15MeV45MeV}} show the exprimental and simulated results of cosmic ray count rate $n^c_k$ as a function of $k$. The highest $n^c_k$ is observed for $k=2$ and decreases by more than one order of magnitude for each additional unit of $k$. The track number analysis for simulations of antiprotons ($\pbar$) irradiating the BGO disk uniformly  with the typical count rate of the $\Hbar$ atoms in the experiment $I_{\pbar}=0.1$~Hz results in the chain curve ($n^{\pbar}_k$).

In this simulation, the CHIPS model was used in Geant4 for $\pbar$ annihilation. This model was previously tested with respect to the energy deposition analysis for $\Hbar$ annihilation in the BGO \cite{ASACUSA,YN1}. The multiplicity of annihilation products from $\pbar$ annihilation was studied using an emulsion detector and agreed with CHIPS results except for annihilation with heavy atoms
\footnote{
It is noted that there are no systematic studies of both multiplicity of annihilation products, and their energy deposition for $\pbar$ annihilation at rest. To investigate this, fragmentation studies of antiproton-nucleus annihilation are being performed using a Timepix3 detector within ASACUSA collaboration.
} \cite{TA}.

The distribution of the chain curve line shows a maximum again, but it decreases weakly as a function of $k$.
When a $\pbar$ annihilates with a nucleus, approximately 3 charged and 2 neutral pions are produced on average \cite{Hori}.
Taking into account the charged pions and the solid angle $\omega$ covered by the hodoscope, $n^{\pbar}_2$ is  estimated to be $3\omega^2(1-\omega)I_{\pbar} \sim 0.04$~Hz which can be compared to $n^{\pbar}_2 \sim 0.03$~Hz in Fig.~{\ref{fig:Count15MeV45MeV}}.

As will be discussed in subsection 5.4, $n^c_k$ can be decreased considerably from the dashed line 
and is around $5 \times 10^{-3}$~Hz as shown by the open circles in Fig.~{\ref{fig:Count15MeV45MeV}} for $k=1$, 2 and 3.

Alternatively, $n^{\pbar}_k$ does not decrease very much as seen from the open triangles.
In these cases, $n^c_k$ is well below $n^{\pbar}_k$.
Further to this, $n^c_4$ is more than one order of magnitude lower than the open circles and is negligibly small.
Therefore we can assume that events for $k \geq 4$  can be reasonably attributed to $\pbar$ annihilation.
In the following subsections, the events for $k=2$, $k=3$ and then $k=1$ are considered.

\begin{figure}[t]
\includegraphics[width=.5\linewidth]{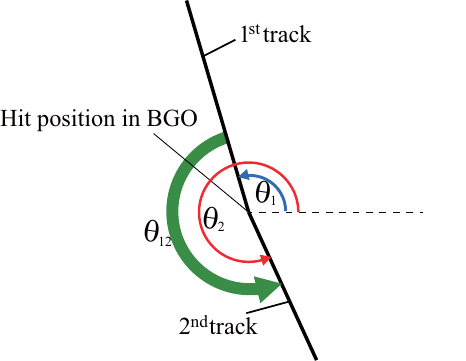}
\caption{\label{fig:tracktheta2}
Definitions of $\theta_1$, $\theta_2$ and $\theta_{12}$.
}
\end{figure}

\subsection{\label{sec:level51}Events for $k=2$}

\begin{figure}[h]
\includegraphics[width=1.\linewidth]{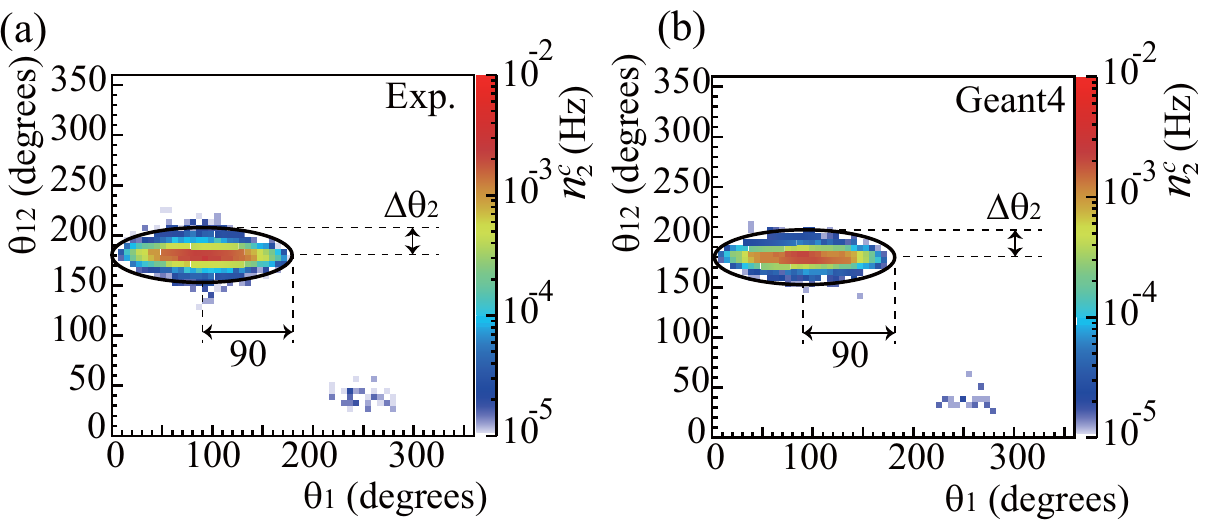}
\caption{\label{fig:AngleDependenceCosmicNt2}
(a) Experimental result of the 2D distribution of $n^c_2$ as a function of $\theta_1$ and $\theta_{12}$.
(b) Simulation result of $n^c_2$ as a function of $\theta_1$ and $\theta_{12}$.
In both figures, $E_{th}=15$~MeV.
The bin widths are 6 degrees on both axes.
The ellipse is defined by semi-minor axis $\Delta \theta_2$ and semi-major axis of 90 degrees
in $\theta_{12}$ and $\theta_1$, respectively, and is used for background suppression.
}
\end{figure}

\begin{figure}[h]
\includegraphics[width=.9\linewidth]{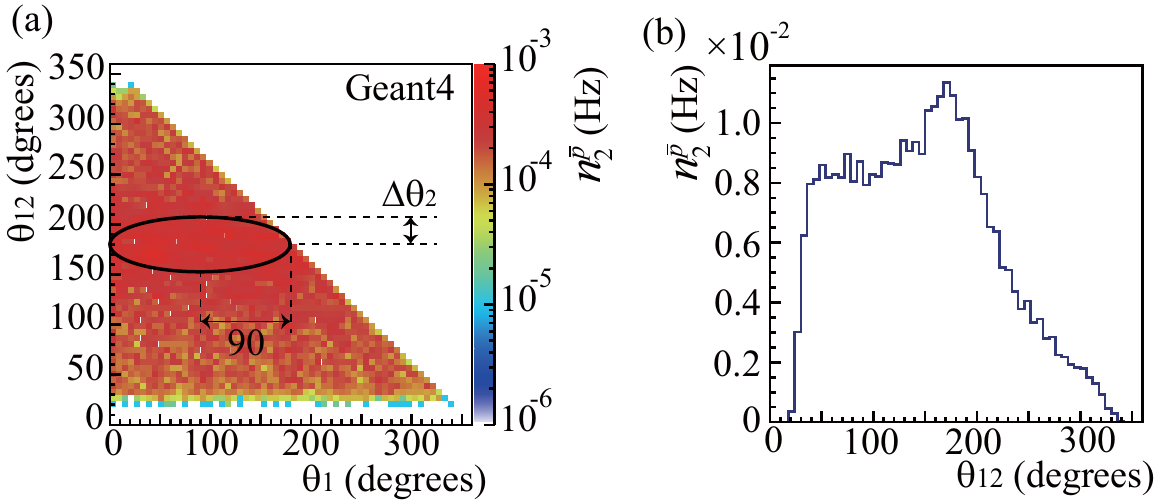}
\caption{\label{fig:AngleDependencePbarNt2}
(a) Experimental result of 2D distribution of $n^{\pbar}_2$ as a function of $\theta_1$ and $\theta_{12}$
with $E_{th}=15$~MeV.
The bin widths are 6 degrees on both axes.
(b) Projection of (a) onto $\theta_{12}$ axis.
The ellipse is defined like in Fig.~{\ref{fig:AngleDependenceCosmicNt2}}.
}
\end{figure}

To analyze 2-track events, the track direction is defined by the angle measured anticlockwise from a horizontal line on the $x-y$ plane as shown in Fig.~{\ref{fig:tracktheta2}}. The tracks are numbered in ascending ordered with increasing angle. The corresponding angles of the $1^{\mathrm{st}}$ and the $2^{\mathrm{nd}}$ tracks are named $\theta_1$ and $\theta_2$ ($\theta_1 < \theta_2$), respectively.
Further we define $\theta_{12} = \theta_2 - \theta_1$.

Figures~{\ref{fig:AngleDependenceCosmicNt2}}~(a) and (b) compare the 2D distribution of cosmic events as a function of $\theta_1$ and $\theta_{12}$ as obtained from experiment and simulation, respectively. 

It can be seen that the simulation result reproduces the experimental result well. A strong ridge is observed at $\theta_{12} \sim 180$~degrees which corresponds to cosmic rays passing straight through the detector.
On the other hand, in $\theta_1$ direction, the distribution spread widely, centered at 90~degrees, which reflects the cosmic ray flux following a $\cos{\theta_z^2}$ distribution.

Figure~{\ref{fig:AngleDependencePbarNt2}}~(a) shows the result of simulation of the 2D distribution of $n^{\pbar}_2$ as a function of $\theta_1$ and $\theta_{12}$. The distribution is very broad, as is expected from $\pbar$ annihilations at low energy.

The $n^c_2$ background can be decreased by removing events inside an ellipse defined by semi-minor axis $\Delta \theta_2$ and semi-major axis of 90 degrees in $\theta_{12}$ and $\theta_1$, respectively (see Fig.~{\ref{fig:AngleDependenceCosmicNt2}}). For example, the $n^c_2$ background is reduced by one order of magnitude for $\Delta \theta_2 = 10$~degrees. Using the same cut, only about 10~\% of the $\pbar$s are removed because of the different distributions
of Fig.~{\ref{fig:AngleDependenceCosmicNt2}} and Fig.~{\ref{fig:AngleDependencePbarNt2}}.

It is noted that in both Figs.~{\ref{fig:AngleDependenceCosmicNt2}}~(a) and (b),
we observe an additional small peak at $\theta_1 \sim 270$~degrees and $\theta_{12} \sim 40$~degrees.
Investigating the corresponding events in the simulation, it was found that energetic $\gamma$ rays in the BGO produce electron-positron pairs and form this peak. The fraction of these events is 1\% of the total events in Fig.~{\ref{fig:AngleDependenceCosmicNt2}}~(a) and (b).

In comparison, although the distribution of $n^{\pbar}_2$ in Fig.~{\ref{fig:AngleDependencePbarNt2}}~(a) is very broad, it has a peak at $\theta_{12} \sim 180$~degrees as shown in Fig.~{\ref{fig:AngleDependencePbarNt2}}~(b) which is the projection of (a) onto the $\theta_{12}$ axis.

The preference at 180 degrees can be explained if we consider a specific type of event, one with three charged pions. When two charged pions hit the hodoscope, with the third pion escaping in a direction close to the beam axis, momentum conservation will favour $\theta_{12}$ around 180~degrees.

\subsection{\label{sec:level52}Events for $k=3$}

\begin{figure}[h]
\includegraphics[width=1.\linewidth]{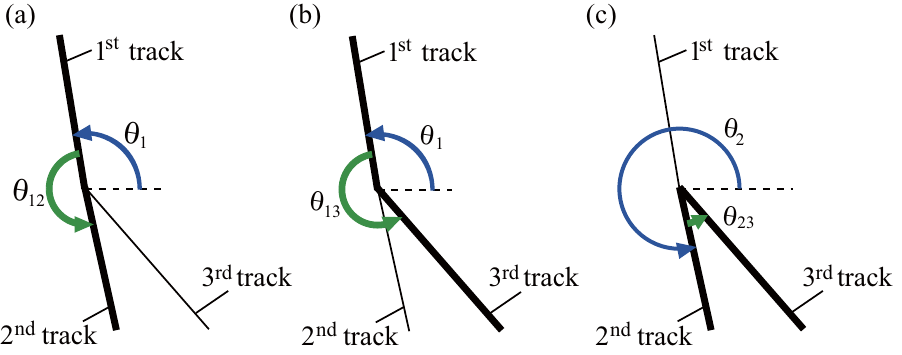}
\caption{\label{fig:track31}
Combinations of 2 out of 3 tracks. Angles $\theta_1$, $\theta_2$, $\theta_{12}$, $\theta_{13}$ and $\theta_{23}$ are defined.
}
\end{figure}

The track directions for $k=3$ are defined in the same manner in the case of $k=2$.
Because 3 tracks are involved, there are 3 ways to choose track pairs, which can then be described in an equivalent manner to 2-track events as shown in Figs.~{\ref{fig:track31}}~(a)-(c).

\begin{figure}[h]
\includegraphics[width=1.\linewidth]{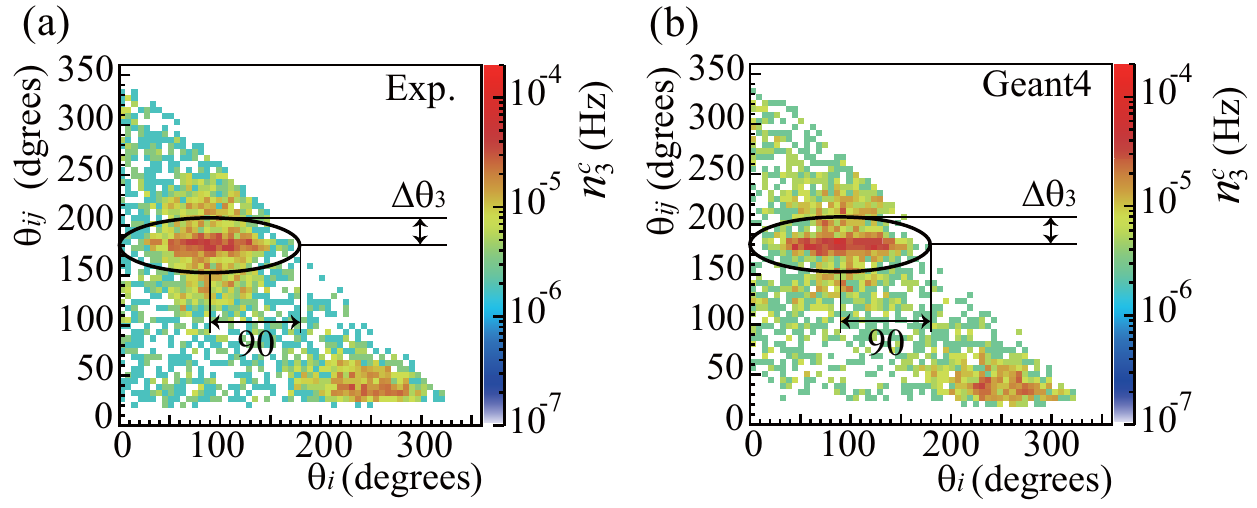}
\caption{\label{fig:AngleDependenceCosmicNt3}
(a) Experimental results of $n^c_3$ obtained by
summing 3 distributions of $\theta_{12}$\,vs.\,$\theta_{1}$, $\theta_{13}$\,vs.\,$\theta_{1}$ and $\theta_{23}$\,vs.\,$\theta_{2}$.
(b) Simulation result of the same distributions as in (a).
In both figures, $E_{th}=15$~MeV.
The bin widths are 6 degrees on both axes.
The ellipse is defined by semi-minor axis $\Delta \theta_3$ and semi-major axis of 90 degrees
in $\theta_{ij}$ and $\theta_i$, respectively, and is used for background suppression.
}
\end{figure}

Figures~{\ref{fig:AngleDependenceCosmicNt3}}~(a) and (b) show the experimental and simulated results of the
2D distributions of $n^c_3$, obtained by summing 3 distributions of $\theta_{12}$\,vs.\,$\theta_{1}$, 
$\theta_{13}$\,vs.\,$\theta_{1}$ and $\theta_{23}$\,vs.\,$\theta_{2}$, i.e. every event is represented by 3 points corresponding to Figs.~{\ref{fig:track31}}~(a)-(c) on the plot to conveniently summarize them.
The simulation reproduces the experiment very well.

A peak at $\theta_{ij} \sim 180$~degrees is observed, which is broader than the peak in Fig.~{\ref{fig:AngleDependenceCosmicNt2}}~(a) and (b). By investigating the corresponding events in the simulation, it was found that a cosmic ray from above generates recoil electrons emitted downward in the BGO which forms the broad peak (see Fig.~{\ref{fig:track31}}~(a) and (b)). Another peak is seen at $\theta_i \sim 270$~degrees
which is formed by the recoil electron together with the incident cosmic ray (see Fig.~{\ref{fig:track31}}~(c)).

\begin{figure}[h]
\includegraphics[width=.5\linewidth]{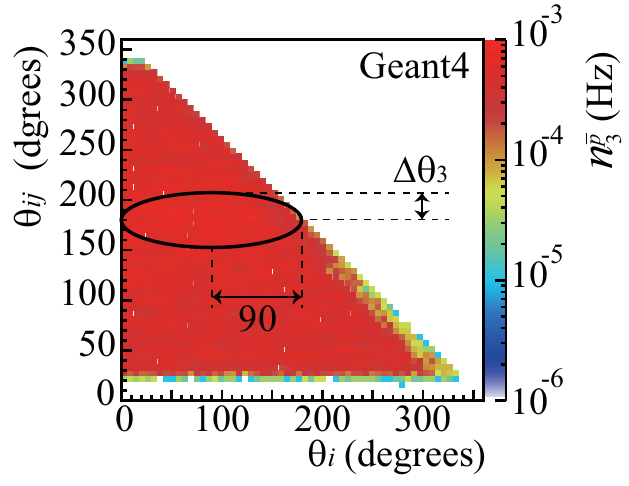}
\caption{\label{fig:AngleDependencePbarNt3}
Result from simulations of $n^{\pbar}_3$ obtained by summing 3 distributions of $\theta_{12}$\,vs.\,$\theta_{1}$, $\theta_{13}$\,vs.\,$\theta_{1}$ and $\theta_{23}$\,vs.\,$\theta_{2}$
with $E_{th}=15$~MeV.
The bin widths are 6 degrees on both axes.
The ellipse is defined as in Fig.~{\ref{fig:AngleDependenceCosmicNt3}}.
}
\end{figure}

Figure~{\ref{fig:AngleDependencePbarNt3}} shows the simulation result of $n^{\pbar}_3$ as per Figs.~{\ref{fig:AngleDependenceCosmicNt3}}~(a) and (b) for $n^c_3$. The event distribution is much broader than $k=2$.

Equivalently to the case of $k=2$,  the cosmic ray events are expected to be reduced by removing events
inside the ellipse (see Fig.~{\ref{fig:AngleDependenceCosmicNt3}}) defined by semi-minor axis $\Delta \theta_3$ and semi-major axis of 90~degrees in $\theta_{ij}$ and $\theta_i$, respectively. Whilst simultaneously not decreasing $n^{\pbar}_3$. It is noted that the peak at $\theta_i \sim 270$~degrees in Figs.~{\ref{fig:AngleDependenceCosmicNt3}}~(a) and (b) is not present when the events inside the ellipse are removed.

\subsection{\label{sec:level53}Signal-to-noise ratio (SNR) for $k \geq 2$}

\begin{figure}[h]
\includegraphics[width=1.\linewidth]{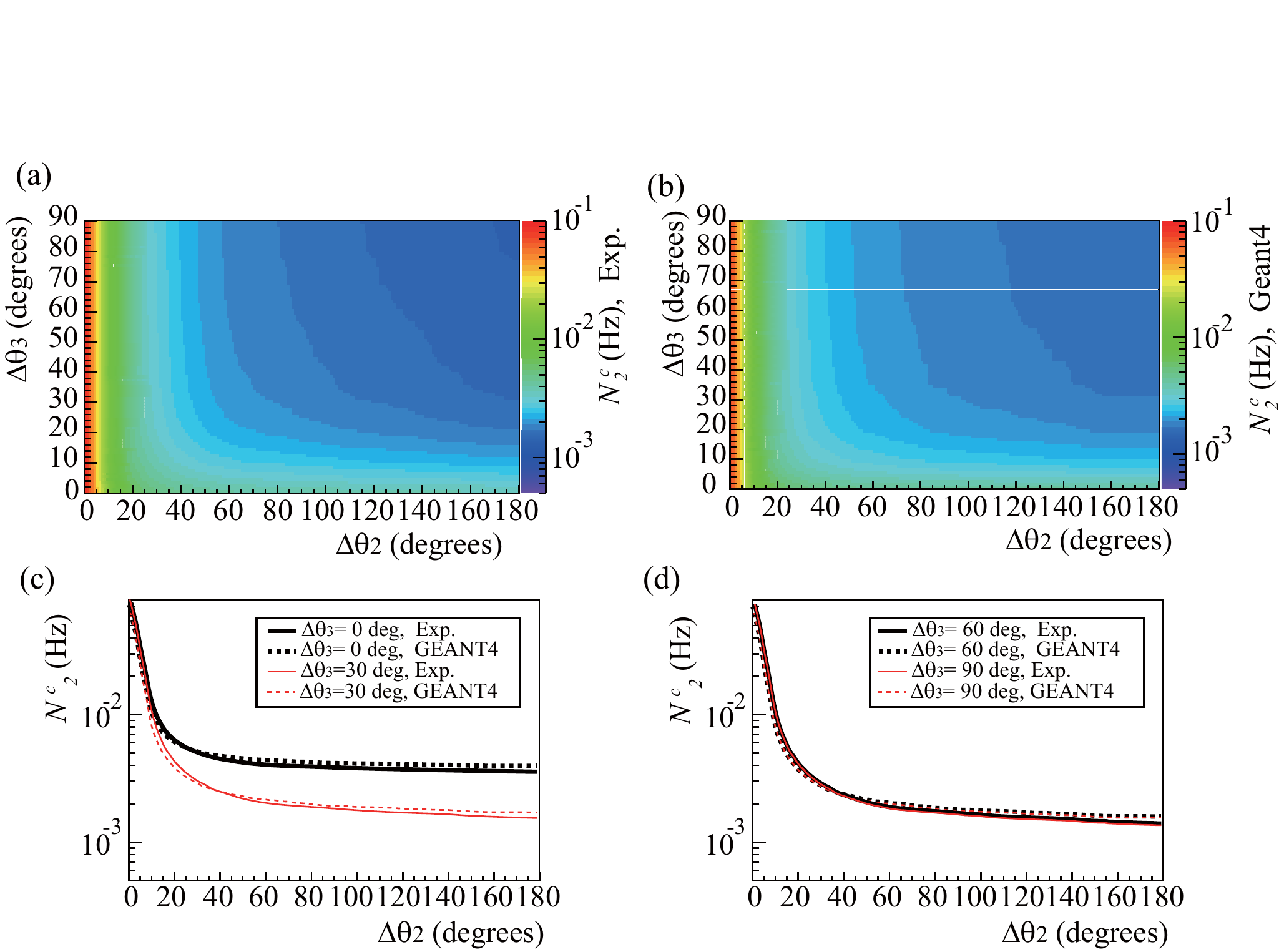}
\caption{\label{fig:significance}
(a) Experimental result of the 2D distribution of $N^c_2$ as a function of $\Delta \theta_2$ and $\Delta \theta_3$.
(b) Corresponding simulation result for $N^c_2$.
The bin widths in Figs.~(a) and (b) are 1 degree on both axes.
(c) and (d) Experimental and simulated results of $N^c_2$ as a function of $\Delta \theta_2$ for $\Delta \theta_3=0$, 30, 60 and 90 degrees.
In these figures, $E_{th}=15$~MeV.
}
\end{figure}

\begin{figure}[h]
\includegraphics[width=1.\linewidth]{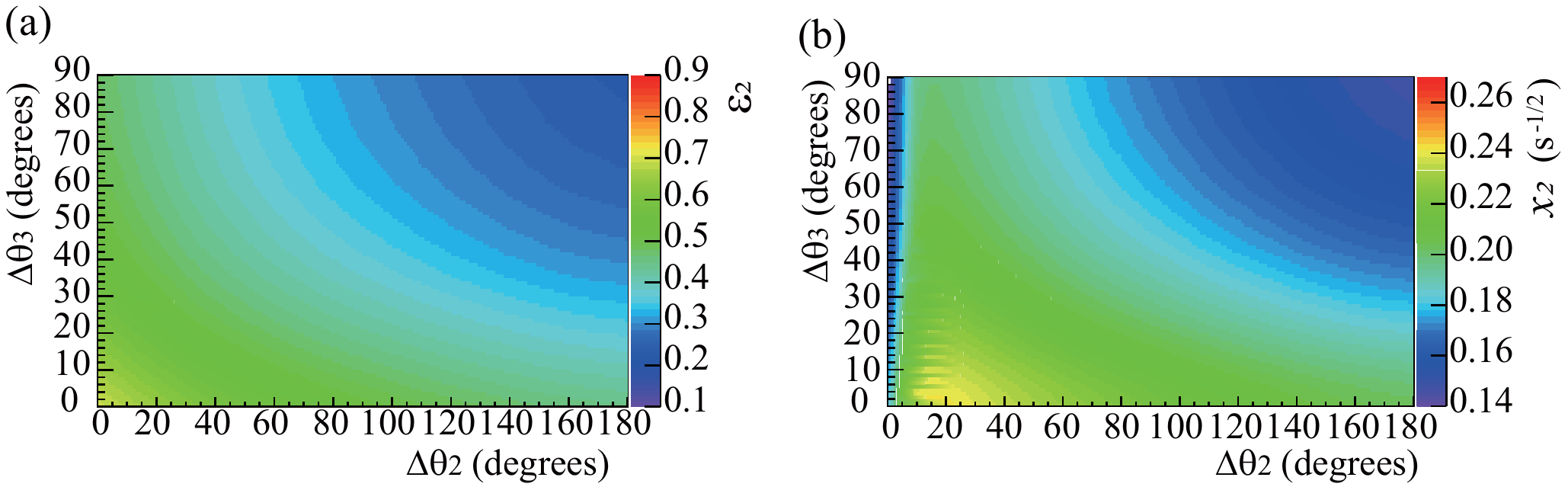}
\caption{\label{fig:significance2}
(a) The simulation result of 2D distribution of $\epsilon_2$ as a function of $\Delta \theta_2$ and $\Delta \theta_3$.
(b) The simulation result of $x_2$ as a function of $\Delta \theta_2$ and $\Delta \theta_3$.
In these figures, $E_{th}=15$~MeV.
The bin widths are 1 degree on both axes.
}	
\end{figure}

\begin{figure}[h]
\includegraphics[width=.6\linewidth]{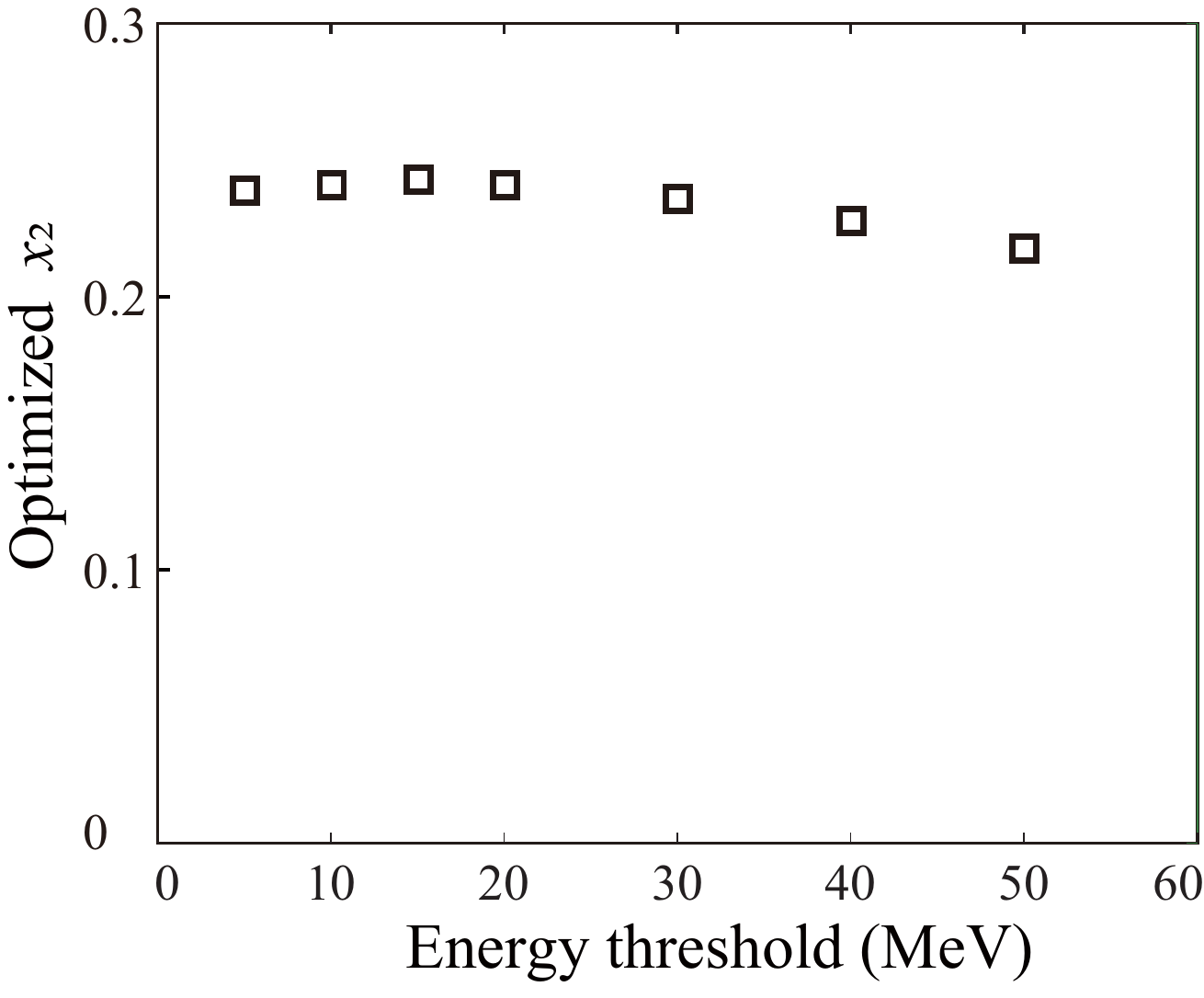}
\caption{\label{fig:snsig}
$x_2$ as a function of $E_{th}$,
where $x_2$ is optimized for each $E_{th}$ using the procedure described in Section~5.3.
}
\end{figure}

The total rate of cosmic events after the data cut was obtained by taking the sum of $n^c_k$ for $k$
and is defined by $N^c_i = \Sigma_{k\geq i} n^c_k$. Figures~{\ref{fig:significance}~(a) and (b) show the experimental and simulated results of the 2D distributions of $N^c_2$, respectively, as a function of $\Delta \theta_2$ and $\Delta \theta_3$.

The simulations reproduce the experimental data well. To evaluate the difference between the data and the simulation quantitatively, Figs~{\ref{fig:significance}~(c) and (d) show experimental and simulated results of $N^c_2$ as a function of $\Delta \theta_2$ for $\Delta \theta_3=0$, $30$, $60$ and $90$~degrees. The difference between the experimental and simulated results is less than 10\%. In the later discussion, we analyze the simulation data.

Figure~{\ref{fig:significance2}~(a) shows the simulation result of the detection efficiency of $\pbar$s $\epsilon_2$ defined by $\epsilon_i = N^{\pbar}_i / I_{\pbar}$ as a function of $\Delta \theta_2$ and $\Delta \theta_3$, where $N^{\pbar}_i=\Sigma_{k\geq i}n^{\pbar}_k$.
It is shown that $\epsilon_2$ decreases as $\Delta \theta_2$ and $\Delta \theta_3$ increase. We define the signal-to-noise ratio (SNR) by $x_i = \frac{N^{\pbar}_i}{\sqrt{N^{\pbar}_i + N^c_i }}$. Figure~{\ref{fig:significance2}~(b) shows the 2D distribution of $x_2$ as a function of $\Delta \theta_2$ and $\Delta \theta_3$. The maximum $x_2$ is 0.24~${\mathrm{s^{-1/2}}}$ at $\Delta \theta_2 = 16$~degrees and $\Delta \theta_3 = 0$~degrees with $N^c_2 = 6.9$~mHz and $\epsilon_2=65$~\%. As is seen in Fig.~{\ref{fig:significance2}~(b), to maximize $x_2$, $\Delta \theta_3$ should be 0 for all values $\Delta \theta_2$. This suggests that all events of $n^{\pbar}_3$ can be identified as $\pbar$s. It is noted that this fact and the optimization of the SNR depend on $I_{\pbar}$.

Figure~{\ref{fig:snsig} shows $x_2$ as a function of $E_{th}$, $x_2$ has a maximum at $E_{th}=15$~MeV, but varying only within 1~\% in the range of 5~MeV~$< E_{th}<20$~MeV. Therefore $E_{th}$ is not critical to the  optimization of $x_2$ in this range.

\subsection{\label{sec:level54}Events for $k=1$ and SNR}

Figure~{\ref{fig:onetrack}~(a) shows the 2D distribution of $n^c_1$ as a function of the deposition energy in the BGO $E$ and $\theta_1$. We observe a ridge at $E \sim 7$~MeV which corresponds to the MIP peak. In the $\theta_1$ direction, the distribution spreads widely with a center at 90 and 270~degrees and its shape is attributed to the cosmic ray flux distribution of $\cos{\theta_z^2}$. The shape of the ridge appears to be an ellipse. However, the tail of the ridge seems to be more a triangular. Figure~{\ref{fig:onetrack}~(b) shows the 2D distribution of $n^{\pbar}_1$ as a function of $E$ and $\theta_1$. The distribution is very broad.
$n^c_1$ is expected to be reduced by removing events inside the triangle defined by $\Delta E$ and the base of 180~degrees as is seen in Fig.~{\ref{fig:onetrack}.

Figures~{\ref{fig:AngleQ}~(a) and (b) show the 2D distributions of $N^c_1$ and $\epsilon_1$ ($k \geq 1$), respectively, as a function of $\Delta \theta_2$ and $\Delta E$. It is seen that $N^c_1$ and $\epsilon_1$ decrease gradually as $\Delta \theta_2$ and $\Delta E$ increase. Figure~{\ref{fig:AngleQ}~(c) shows $x_1$ as a function of $\Delta \theta_2$ and $\Delta E$. The maximum of $x_1$ reached 0.26~${\mathrm{s^{-1/2}}}$ at $\Delta \theta_2 = 14$~degrees and $\Delta E = 93$~MeV  with $N^c_1=12$~mHz and $\epsilon_1=81$~\%.
This is larger than the maximum value of $x_2$, therefore, the analysis including the events for $k=1$ improves the SNR. Comparing with the $\Hbar$ detector developed in 2012 with $x=0.22$~${\mathrm{s^{-1/2}}}$ with the cosmic count rate of $4$~mHz and the detection efficiency of $50$~\% for $I_{\pbar}=0.1$~Hz,
the detector described in this work improves upon the SNR and the detection efficiency.

\begin{figure}[h]
\includegraphics[width=1.\linewidth]{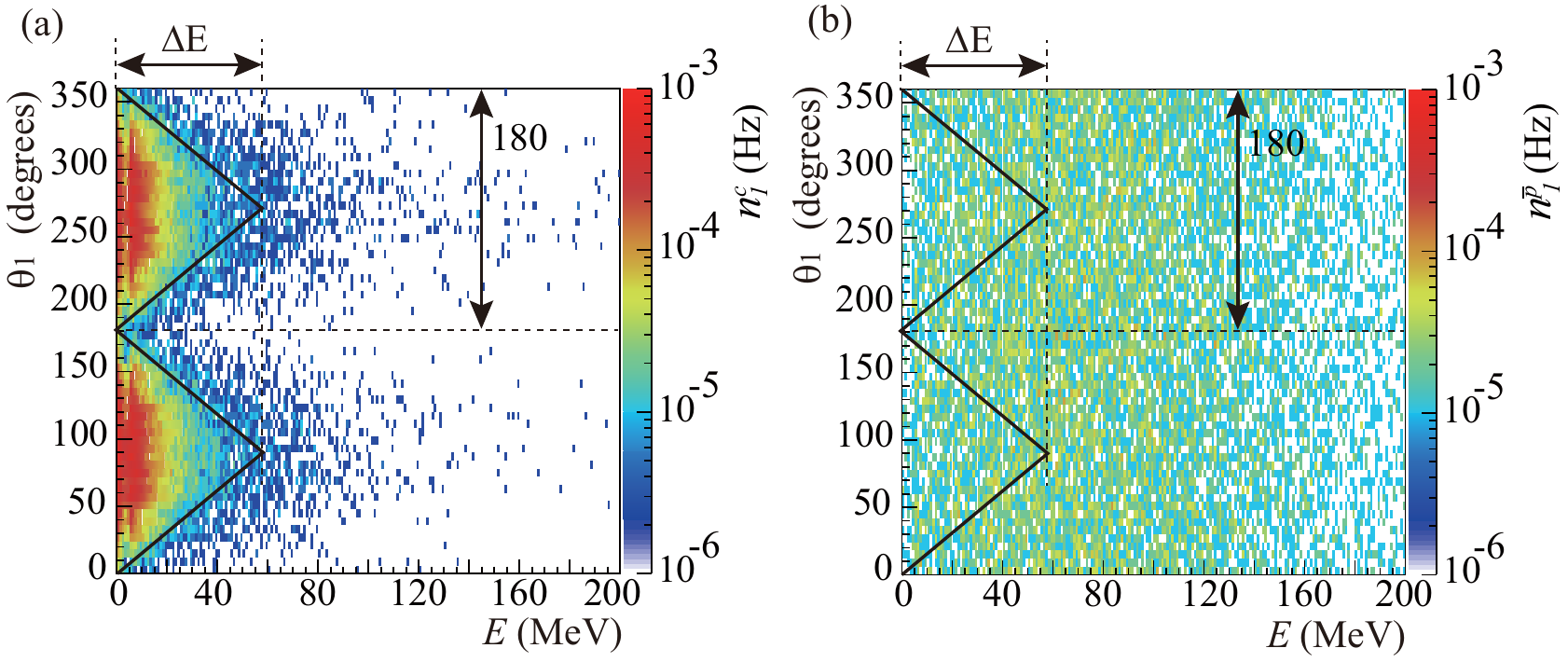}
\caption{\label{fig:onetrack}
(a) Simulation result of the 2D distribution of $n^c_1$ as a function of $E$ and $\theta_1$.
(b) Simulation result of the 2D distribution of $n^{\pbar}_1$ as a function of $E$ and $\theta_1$.
The bin widths in horizontal and vertical axes are 1~MeV and 6~degrees, respectively.
}
\end{figure}

\begin{figure}[h]
\includegraphics[width=1.\linewidth]{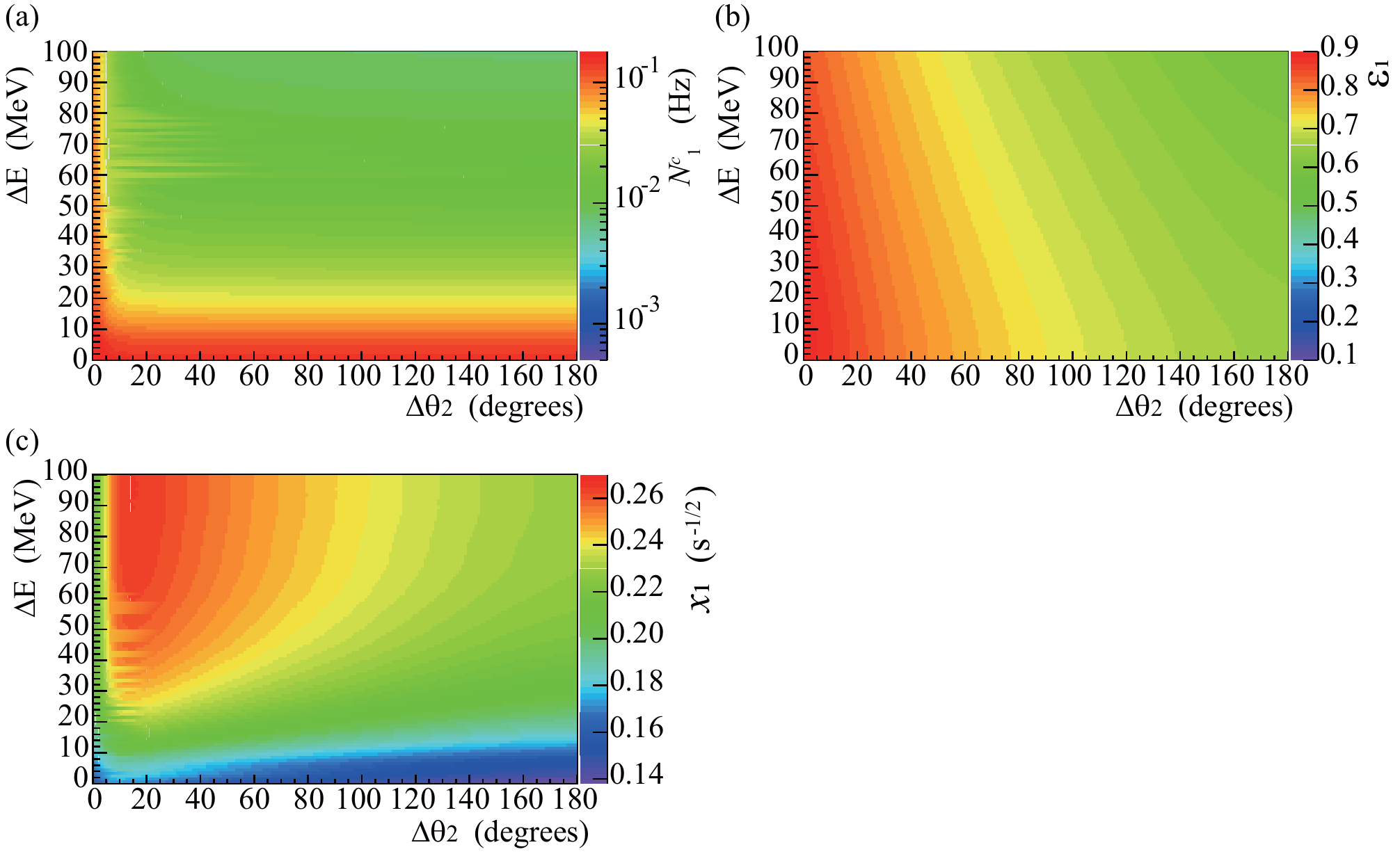}
\caption{\label{fig:AngleQ}
(a) Simulation result of 2D distributions of $N_1^c$ as a function of $\Delta \theta_2$ and $\Delta E$.
(b) $\epsilon_1$ as a function of $\Delta \theta_2$ and $\Delta E$.
(c) $x_1$ as a function of $\Delta \theta_2$ and $\Delta E$.
In these figures, $E_{th}$ for $k \geq 2$ is $15$~MeV.
The bin widths are 1 degree on both axes.
}
\end{figure}

\section{\label{sec:level6}Conclusion}

We have developed a $\Hbar$ detector consisting of a thin BGO disk and a hodoscope. We have measured hit positions of cosmic rays in the BGO disk and confirmed that the thin disk with a 2D readout by MAPMTs enables position sensitivity. The energy deposition in the BGO was calibrated by comparing cosmic ray data with Geant4 simulations. Charged particle tracks were determined by connecting the hit position on the BGO and hits on hodoscope bars. By removing the cosmic rays passing through the detector using the cut on $\Delta E$, $\Delta \theta_2$ and $E_{th}$, the background was reduced efficiently to $N^c_1=12$~mHz with a detection efficiency of $\epsilon_1=81$~\%. The SNR was improved to $x_1=0.26$~${\mathrm{s^{-1/2}}}$ which was compared to 0.22~${\mathrm{s^{-1/2}}}$ for the detector used in 2012.

\section*{Acknowledgements}

We would like to thank Tomohiro Kobayashi for the carbon coating on the BGO disk.
This work was supported by the Grant-in-Aid for Specially Promoted Research 24000008 of Japanese Ministry of Education, Culture, Sports, Science and Technology (MEXT),
Special Research Projects for Basic Science of RIKEN,
European Research Council under European Union's Seventh Framework Programme (FP7/2007-2013)
/ERC Grant Agreement (291242) and the Austrian Ministry of Science and Research, Austrian Science Fund (FWF): W1252-N27.



\begin{thebibliography}{20}
\bibitem{ASACUSA}N. Kuroda et al., A source of antihydrogen for in-flight hyperfine spectroscopy, Nat. Commun. {\bf{5}} (2014) 3089.
\bibitem{MY} A. Mohri, and Y.Yamazaki, A possible new scheme to synthesize antihydrogen and to prepare a polarized antihydrogen beam, Europhys. Lett. {\bf{63}} (2003) 207.
\bibitem{YY} Y. Nagata and Y. Yamazaki, A novel property of anti-Helmholz coils for in-coil syntheses of antihydrogen atoms: formation of a focused spin-polarized beam, New J. Phys. {\bf{16}} (2014) 083026.
\bibitem{DC} Y. Nagata et al., The development of the superconducting double cusp magnet for intense antihydrogen beams, Journal of Physics: Conference Series {\bf{635}} (2015) 022062.
\bibitem{SMI} M. Diermaier et al., In-beam measurement of the hydrogen hyperfine splitting and prospects for antihydrogen spectroscopy, Nat. Commun. {\bf{8}} (2017) 15749.
\bibitem{HFS1} ASACUSA proposal addendum, CERN/SPSC 2005-002, SPSC P-307 Add.1 (2005).
\bibitem{HFS2} E. Widmann et al., Measurement of the hyperfine structure of antihydrogen in a beam, Hyperfine Interact {\bf{215}} (2013) 1.
\bibitem{NK} N. Kuroda et al., Antihydrogen Synthesis in a Double-Cusp Trap,  JPS Conf. Proc. {\bf{18}} (2017) 011009.
\bibitem{MT} M. Tajima et al., Manipulation and transport of antiprotons for an efficient production of antihydrogen atoms, JPS Conf. Proc. {\bf{18}} (2017) 011008.
\bibitem{chloe} C. Malbrunot et al., The ASACUSA antihydrogen and hydrogen program : results and prospects, Phil. Trans. Roy. Soc. A,  {\bf{376}} (2018) 20170273. 
\bibitem{YN1} Y. Nagata et al., Direct detection of antihydrogen atoms using a BGO crystal, Nucl. Instrum. Methods Phys. Res. Sect. A {\bf{840}} (2016) 153. 
\bibitem{CS2} C. Sauerzopf et al., Annihilation detector for an in-beam spectroscopy apparatus to measure the ground state hyperfine splitting of antihydrogen, Nucl. Instrum. Methods Phys. Res. Sect. A, {\bf{845}} (2016) 579.
\bibitem{YN2} Y. Nagata et al., The development of the antihydrogen beam detector: toward the three dimensional tracking with a BGO crystal and a hodoscope, JPS Conf. Proc. {\bf{18}} (2017) 011038.
\bibitem{CS1} C. Sauerzopf et al., Intelligent Front-end Electronics for Silicon photodetectors (IFES), Nucl. Instrum. Methods Phys. Res. Sect. A, {\bf{819}} (2016) 163.
\bibitem{Geant} S.Agostinelli, et al., Geant4.a simulation toolkit, Nucl. Instrum. Methods Phys. Res. Sect. A {\bf{506}} (3) (2003) 250.
\bibitem{Hagmann} C. Hagmann et al., IEEE Nucl. Sci. Symp. Conf. Record 2 (2007) 1143.
\bibitem{leo} W. R. Leo, Techniques for Nuclear and Particle Physics Experiments, Springer-Verlag (1993).
\bibitem{PDG} K.A. Olive et al. (Particle Data Group), Chin. Phys. C, {\bf{38}} (2014) 090001.
\bibitem{TA} S. Aghion et al., Measurement of antiproton annihilation on Cu, Ag and Au with emulsion films, JINST {\bf{12}} (2017) P04021.
\bibitem{Hori} M. Hori, et al., Analog Cherenkov detectors used in laser spectroscopy experiments on antiprotonic helium, Nucl. Instrum. Methods Phys. Res. Sect. A {\bf{496}} (2003) 102.
\end{thebibliography}
\end{document}